\documentclass[%
reprint,
amsmath,amssymb,
aps,
pra,
]{revtex4-1}

\usepackage{appendix}
\usepackage{graphicx}
\usepackage{dcolumn}
\usepackage{bm}

\usepackage{subfigure}

\begin{document}


\title{Neural network state estimation for full quantum state tomography}

\author{Qian Xu}
\thanks{These authors contributed equally to this work.}
\affiliation{School of Physics, Nanjing University, Nanjing 210093, China}

\author{Shuqi Xu}
\thanks{These authors contributed equally to this work.}
\affiliation{School of Physics, Nanjing University, Nanjing 210093, China}


\date{\today}

\begin{abstract}
An efficient state estimation model, neural network estimation (NNE), empowered by machine learning techniques, is presented for full quantum state tomography (FQST). A parameterized function based on neural network is applied to map the measurement outcomes to the estimated quantum states. Parameters are updated with supervised learning procedures. From the computational complexity perspective our algorithm is the most efficient one among existing state estimation algorithms for full quantum state tomography. We perform numerical tests to prove both the accuracy and scalability of our model.
\end{abstract}

\pacs{Valid PACS appear here}
\maketitle


\section{\label{sec:level1}Introduction}
Quantum state tomography (QST) \cite{lvovsky2009continuous}, reconstructing quantum state of a quantum system via quantum measurements, plays an important role in verifying and benchmarking quantum devices in various quantum information processing assignments, including quantum computation  \cite{nielsen2002quantum} and quantum communication \cite{mattle1996dense}. QST is comprised of two processes in general: quantum measurements on the system (data collection) and the estimation of the quantum state from the measurement results (data analysis). Both processes are nontrivial.

Up to now, plenty of methods have been proposed to reduce the measurements in demand. For example, some use compressed sensing to reconstruct quantum states with low rank \cite{gross2010quantum, flammia2012quantum, smith2013quantum}; some use efficient tomography methods \cite{baumgratz2013scalable, cramer2010efficient} to reconstruct matrix product state; some use permutationally invariant tomography \cite{toth2010permutationally} to obtain information which is constant under permutation; some can directly estimate the purity of the prepared state \cite{bagan2005purity} and the fidelity compared with the ideal state \cite{flammia2011direct}; Some use generative network (restricted Boltzmann machine) to construct variational representation of many-body quantum system and reconstruct the many-body wavefunction of some certain quantum system with limited measurement results \cite{Torlai2018direct}. However, most of them either can only be implemented under certain prerequisites, such as a prior knowledge or assumptions, or have significant limitations. 

Full quantum state tomography (FQST), which refers to reconstructing quantum state using informationally complete measurements and requires no prior knowledge about the system, is the most versatile tomographic scheme to date. To reconstruct a d-dimensional density matrix, one needs to determine $(d^2-1)$ independent parameters, which means at least $(d^2-1)$ measurements are required. As d grows exponentially with the system size, i.e. the number of qubits, the measurements and computational complexity grow exponentially as well. In additional to the increase of measurements, the existing state reconstruction algorithms are also time consuming. For example eight-qubit reconstruction with maximum likelihood estimation (MLE) \cite{d20042} took 10 hours in measurements \cite{da2011practical, haffner2005scalable} while a week in data processing. Bayesian mean estimation (BME) \cite{blume2010r}, another commonly-used state estimation algorithm gives a unique state estimate but has larger computational complexity. Since the number of qubits in controllable quantum systems is still in rapid growth, more efficient state reconstruction methods are in urgent need. Recently an efficient MLE algorithm has been proposed for state reconstruction from measurements with additive Gaussian noise, which has a complexity of $O(d^4)$, although it is not general \cite{smolin2012efficient}. Besides, a recently-presented linear regression estimation (LRE) algorithm has $O(d^4)$ computational complexity \cite{qi2013quantum} and the computational complexity of its accelerated version is $O(d^{\log_{2}12})$ \cite{hou2016full}. 

Machine learning (ML) techniques have been proven to be powerful tools for recognizing, classifying, and characterizing complex sets of data. In additional to its computer science applications, ML has recently been used to address problems in Physics. For example, its application in various condensed matter topics is an emerging and burgeoning field \cite{curtarolo2003predicting, snyder2012jc, hautier2010finding}. Also, due to its intrinsic capability of exacting information from high-dimensional data, its applications in quantum many-body systems, such as discovering phase transitions, \cite{wang2016discovering, carrasquilla2017machine, broecker2017machine, van2017learning}, solving quantum impurity problems \cite{arsenault2014machine}, representing many-body quantum states \cite{carleo2017solving}, have been paid great attention.

In this paper, we propose an efficient state estimation model, neural network estimation (NNE), empowered by machine learning techniques, to further speed up the data processing of full quantum state tomography. The computational complexity is $O(d^3)$, which is even faster than LRE method.  We turn the state estimation into a regression process and apply a parameterized function based on a neural network structure to map the measurement outcomes onto the estimated quantum states. Standard supervised learning procedures are applied to update the parameters of the parameterized function and numerical tests are performed to prove both the accuracy and scalability of our model.

Our paper is structured as follows: In Sec.II we give detailed description of our model including the structure of the parameterized function we use, generation of the training set and the training process. In Sec.III we perform numerical error evaluations on our model. In Sec.IV we analyze the computational complexity of our model. Sec.V is discussion and conclusion. 

\section{\label{sec:level1}Model}
Here we propose an efficient state estimation model based on supervised learning in ML. According to the universal approximation theorem \cite{le2008representational}, a multi-layer full-connected neural network with proper number of neurons in each layer can approximate any continuous function on the compact subsets of $R^n$. By constructing a mapping function between the measurement results and the reconstructed quantum state based on neural network and turning the estimation task to a regression problem we can use standard supervised learning procedure to finish the task. 

The measurement results, in general, are represented by a high-dimensional feature vector. This feature vector, especially its dimension, depends on the choice of measurement sets. Many results have been presented for choosing optimal measurement sets to increase the estimation accuracy or efficiency in quantum state tomography \cite{adamson2010improving, wootters1989optimal, de2008choice}. There are also many discussions about how to reduce the number of measurements. However, the minimum number of measurements required in the versatile full quantum state tomography task is $d^2$, where $d$ denotes the dimension of the Hilbert space. Here we choose Pauli measurements with $6^n$-dimensional feature vector, which is informationally overcomplete. There are also several ways to represent the reconstructed quantum state. Here we choose the density matrix, which is the most general one to describe both pure states and mixed states.

The main reconstruction process of our model is presented as follows: 

Firstly, we construct a highly parameterized function, which includes a neural network structure to improve its expressive capability, to map a high-dimensional vector comprising of measurement outcomes to the density matrix of the state to be reconstructed. Then we train our model by optimizing parameters based on error back propagation on a large training set. Each training example in the training set is a pair constituting of a vector of measurement outcomes  (input feature) and a corresponding density matrix (label). After the network is well trained, i.e. all the parameters are set properly, one can get the estimated density matrix of the quantum state efficiently from new experiment measurement results through a single feed-forward process. Details are presented in the following sections.

\subsection{Structure of the parameterized function}
\begin{figure}[h!]
    \centering
    \includegraphics[height = 0.4\textwidth,width = 0.4\textwidth]{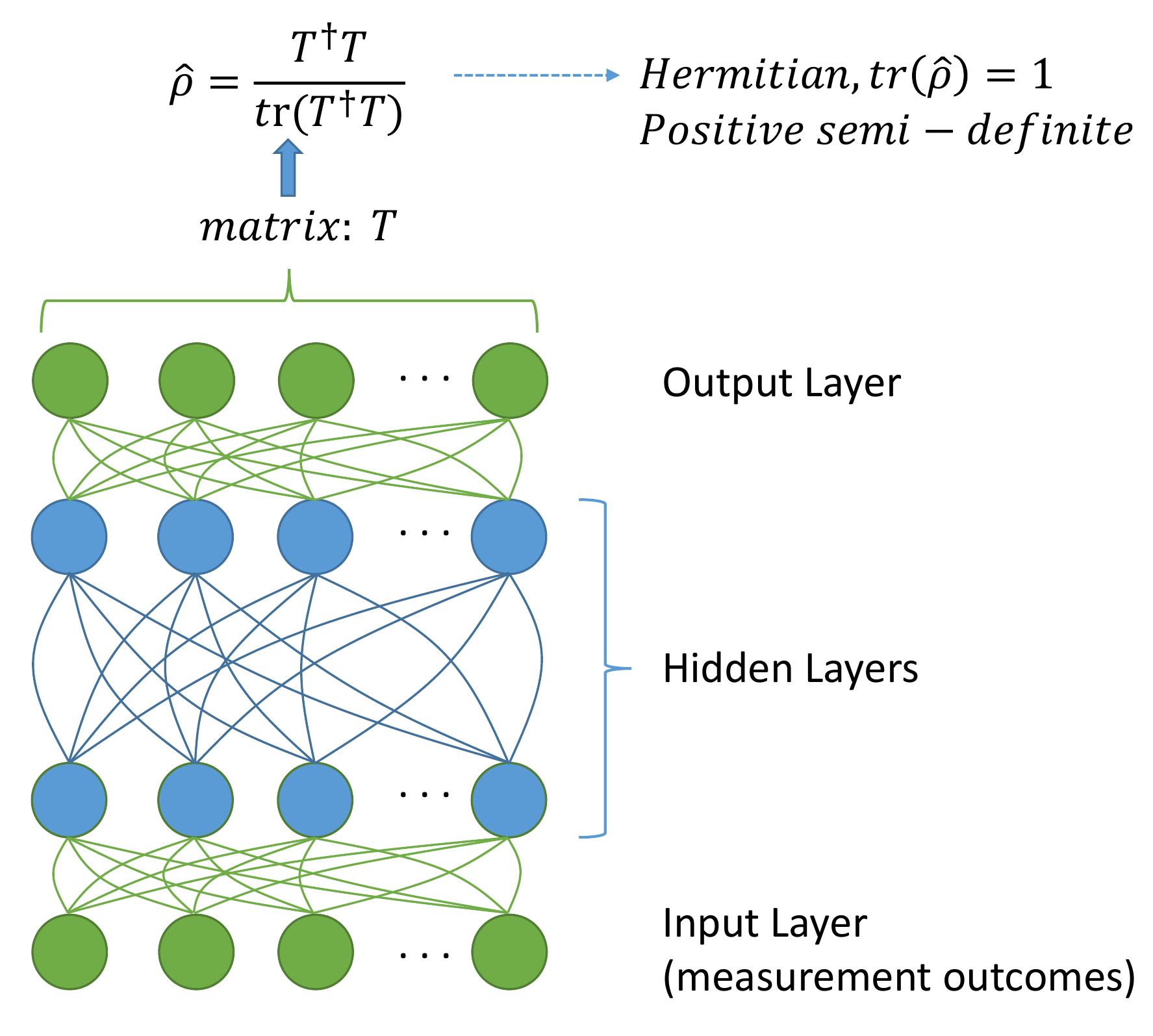}
    \caption{Structure of the parameterized function in our model.}
    \label{fig:structure}
\end{figure}

As Fig.~\ref{fig:structure} shows, the feed-forward mapping from the feature measurement vector to the output density matrix comprises of two steps.

Step(i). A matrix $T$ is mapped from a feature vector $\vec{\alpha}$ by the feed-forward calculation through a four-layer fully-connected neural network \cite{goodfellow2016deep}. The first layer is the input layer. The number of neurons in this layer is the dimension of the input feature vector. For example, there will be $6^n$ neurons in the first layer if Pauli measurements are chosen. The number of neurons in each of the two hidden layers is chosen to linearly increase with the number of qubit number $n$. In practice, we set 200 neurons in each hidden layer for one-qubit system and for 7-qubit system it increases to 800. The matrix T is output from the output layer in the vector form (We separate the real and imaginary part of T for convenience). As the density matrix is $2^n*2^n$, there are $2*4^n$ neurons in the output layer. All neurons are activated by the sigmoid function. (Different from some deep learning tasks like image recognition, in which relu activation function is more preferable, we find sigmoid function is more efficient in our task).

Step(ii). The state represented by density matrix $T$ is not physical yet. A density matrix representing a quantum state should satisfy three conditions: Hermitian, positive semi-definite and of trace one. In order to pull T back to a physical state, we set

\begin{equation}
    \hat{\rho} = \frac{T^\dagger T}{\rm{Tr}(T^\dagger T)}
\end{equation}

Then the state estimate $\hat{\rho}$ satisfies all the three physical constraints.



\subsection{Generation of the training set}

The training set consists of feature vectors and training targets. Training targets are a set of density matrices under a certain distribution and feature vectors are the corresponding ideal measurement outcomes for a given measurement set. 

First we generate the set of training targets. An ensemble of general states of a quantum system is in general described by a probability prior. For our training purpose, this probability prior should be able to produce a uniform spread throughout the state space. In our case this uniformity is defined based on the Bures distance and an ensemble of training targets (density matrices) $\{\rho_t\}$ are generated according to the Bures distribution \cite{zyczkowski2001induced, hall1998random, pozniak1998composed} (See Appendix A for details).

Then we can generate the feature vectors corresponding to the training targets for a given measurement set $\{\hat{M}\}$. For each $\rho_{t_i}$, its corresponding feature vector consists of the ideal measurement outcomes:
\begin{equation}
\vec{v}_i = ({\rm Tr}(\hat{M}_1 \rho_{t_i}), {\rm Tr}(\hat{M}_2 \rho_{t_i}) ... {\rm Tr}(\hat{M}_N \rho_{t_i}))^T
\end{equation}
where $N$ denotes the total number of measurements. 
For Pauli measurements, the projective measurement set for 1-qubit system is:
\begin{equation}
    \{\hat{M}\} = \{|H\rangle \langle H|, |V\rangle \langle V|, |D\rangle \langle D|, |A\rangle \langle A|, |R\rangle \langle R|, |L\rangle \langle L|\}
\label{eq:1-qubit_measurement}
\end{equation}
where
$$ |H\rangle = \begin{pmatrix}
   1 \\
   0\\
  \end{pmatrix};\ \ |V\rangle = \begin{pmatrix}
   0 \\
   1\\
  \end{pmatrix} $$
  
$$ |D\rangle = \frac{1}{\sqrt{2}}\begin{pmatrix}
   1 \\
   1\\
  \end{pmatrix};\ \ |A\rangle = \frac{1}{\sqrt{2}}\begin{pmatrix}
   1 \\
   -1\\
  \end{pmatrix}; $$
$$ |R\rangle = \frac{1}{\sqrt{2}}\begin{pmatrix}
   1 \\
   i\\
  \end{pmatrix};\ \ |L\rangle = \frac{1}{\sqrt{2}}\begin{pmatrix}
   1 \\
   -i\\
  \end{pmatrix} $$

For the multi-qubit system the measurement set $\{M\}$ is comprised of tensor products among elements in Eq.~\ref{eq:1-qubit_measurement}.

In this paper we use the Pauli measurement set, which is informationally overcomplete, to generate the training set due to its generality and high level of estimation accuracy.

\subsection{Training of the model - supervised learning}
As we have adapted the estimation task to a standard supervised learning task with neural network in ML, we use tensorflow framework in python to run our model. We train the network with stochastic gradient descent algorithm. The optimization goal is the least-square error (LSE) between the state estimate $\hat{\rho}$ (see Fig.~\ref{fig:structure}) and the target matrix $\rho_{t}$ in the training set.
\begin{equation}
    LSE(\hat{\rho},\rho_{t}) = \sqrt{\frac{\sum_{i=1}^{2^n} \sum_{j=1}^{2^n}|M_{ij} - \rho_{t_{ij}}|^2}{4^n}}
\end{equation}
We choose AdamOptimizer as the optimizer to minimize the cost function for its capability to jump out of the local minima \cite{kingma2014adam}. The typical batch size is 200.

\begin{figure}[h!]
    \centering
    \includegraphics[height = 0.4\textwidth,width = 0.5\textwidth]{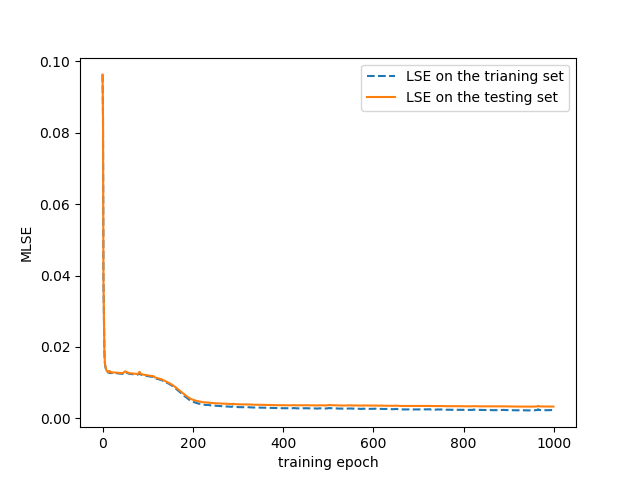}
    \caption{Training process of a 2-qubit network. The yellow line is the MLSE on the testing set and the blue dotted line is the MLSE on the training set. The training is ceased when the MLSE on the testing set stops decreasing.}
    \label{fig:training_curve}
\end{figure}

Because in the training set the measurement outcomes are all ideal without noise while in practice measurement outcomes contain noise, there will be overfitting problem. To solve this, we utilize the testing set generated from simulated experiments. In our training process we use the testing set with $N_0 = 1000$, which is described in detail in the next subsection. Training is terminated when the mean least-square error (MLSE) over the testing set converges. Fig.~\ref{fig:training_curve} shows the training process for a 2-qubit system.

We have trained our network for quantum system with up to 5 qubits. Table,~\ref{tab:training_parameters} shows the number of neurons in two hidden layers and the number of training examples we use for each network. These parameters are what we have currently found to obtain best results. It is worth mentioning that the more training examples, the smaller least-square error the network will reach, i.e. the more accurate this model will be. We use limited training examples because of limited computation resources and time. Evaluation results in the next section are based on networks trained with these training parameters.

\begin{table}[h!]
    \centering
    \begin{tabular}{p{2cm}<{\centering}|p{3cm}<{\centering}|p{3cm}<{\centering}}
         \hline
         Number of qubit & Number of neurons in two hidden layers & Number of training examples\\
         \hline
         1 & 200; 200 & 10000 \\
         \hline
         2 & 300; 300 & 20000 \\
         \hline
         3 & 400; 400 & 40000 \\
         \hline
         4 & 600; 600 & 80000 \\
         \hline
         5 & 800; 800 & 100000 \\
         \hline
    \end{tabular}
    \caption{The number of neurons in two hidden layers and the number of training examples of networks for system with different number of qubits.}
    \label{tab:training_parameters}
\end{table}


\section{\label{sec:level1}Error evaluation}
\subsection{Evaluation by numerical simulations of tomography}

We use numerical simulations of tomography \cite{de2008choice} to evaluate the accuracy of our model. We briefly describe the process here:

\begin{enumerate}
  \item Choose a testing set $\{\rho_i\}$ with $N_s$ states.
  \item For each $\rho_i$, simulate measurements on $N_c$ copies of the state (divided equally between the measurement sets, for each measurement set there are $N_0$ copies of the state) generating $N_c$ measurement outcomes. For our simulations, we chose Pauli measurement set.
  \item Perform state estimation using our model based on these $N_c$ measurement outcomes to obtain the state estimate $\hat{\rho}_i$.
  \item For each pair of state estimate and true state, calculate the point fidelity $f(\hat{\rho}_i(N_0),\rho_i)$：
    \begin{equation}
        f(\hat{\rho}_i,\rho_i) = [{\rm Tr}(\sqrt{\sqrt{\hat{\rho}_i}\rho\sqrt{\hat{\rho}_i}})]^2
    \label{eq:point_fidelity}
    \end{equation}
    
    \item Calculate the average fidelity over the testing state set.
\begin{equation}
    F_{av}(N_0) = \sum_{i=1}^{N_s}f(\hat{\rho}_i,\rho_i)/N_s
\label{eq:average fidelity}
\end{equation}
\end{enumerate}

\begin{figure}[h!]
    \centering
    \includegraphics[width = 0.4\textwidth]{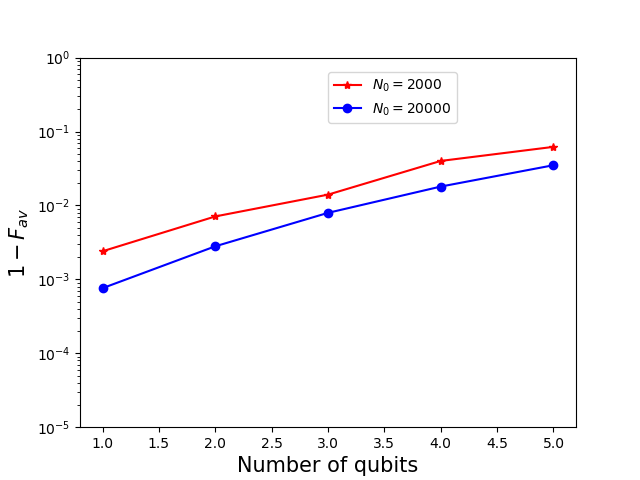}
    \caption{The average fidelity of NNE on a testing set with Bures prior.}
    \label{fig:F-n}
\end{figure}

First to test the overall performance of our model on different states we generate the testing set with $5000*2^n$ states according to the Bures prior. $n$ denotes the number of qubits. Then we perform numerical simulations and calculate the average fidelity for system with different numbers of qubits. The result is shown in Fig.~\ref{fig:F-n}. Evaluations at $N_0 = 20000$ and $N_0 = 2000$ are conducted. We note that $N_0$ reflects the strength of Gaussian noise in the testing data resulting from limited sampling number. 

Then we compare the performance of NNE and MLE. Because MLE is time consuming it is not realistic to calculate its average fidelity on the testing set with Bures prior, especially when the number of qubits in the system is large. So we compare their performance on two kinds of specific and representative states:

\begin{figure}[h!]
\centering
\subfigure{
 \includegraphics[width = 0.45\textwidth ]{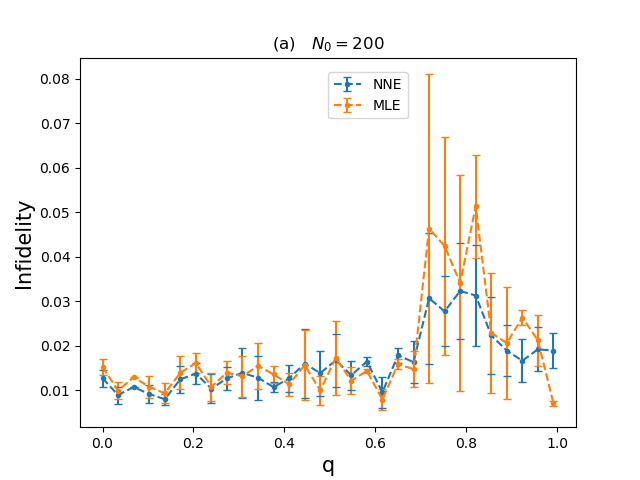}
 \label{fig:Werner_N0_200}
 }
 \hspace{0.1in}
 \subfigure{
 \includegraphics[width = 0.45\textwidth ]{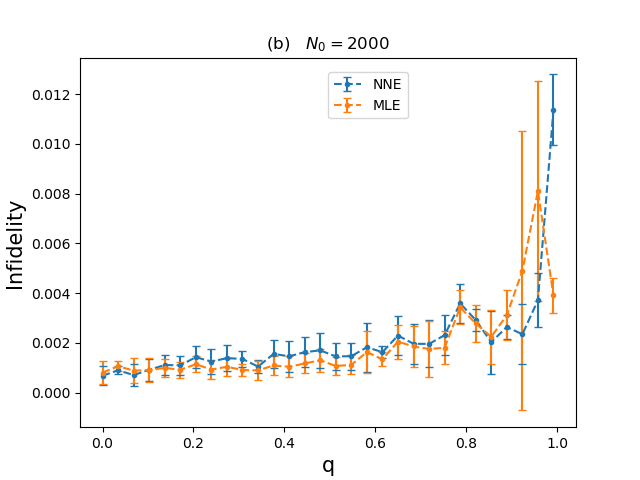}
 \label{fig:Wernere_n0_20000}
 }
 \caption{Infidelity of NNE and MLE on Werner states. (a) $N_0 = 200$. (b) $N_0 = 2000$}
 \label{fig:Werner_state}
\end{figure}

The first state is Werner state:
\begin{equation}
    \rho_{\rm Werner} = q|\Psi^-\rangle \langle\Psi^-| + \frac{1-q}{4}I
\label{equ:Werner}
\end{equation}
where $|\Psi^-\rangle = \frac{|HV\rangle - |VH\rangle}{\sqrt{2}}$

By varying $q$ the state transits from maximally entangled state to maximally mixed state. Fig.~\ref{fig:Werner_state} shows the comparison between NNE and MLE on Werner state.

The infidelity between two states $\rho_i$ and $\rho_j$ is defined as:
\begin{equation}
    {\rm Infidelity}(\rho_i,\rho_j) = 1 - f(\rho_i,\rho_j)
\end{equation}
where $f$ is fidelity.

We can see our model has similar performance compared to MLE on Werner states.

The second specific state class is the maximally mixed state:
\begin{equation}
    \rho_{\rm test} = \frac{I}{2^n}
\end{equation}

\begin{figure}[h]
    \centering
    \includegraphics[ width = 0.4\textwidth]{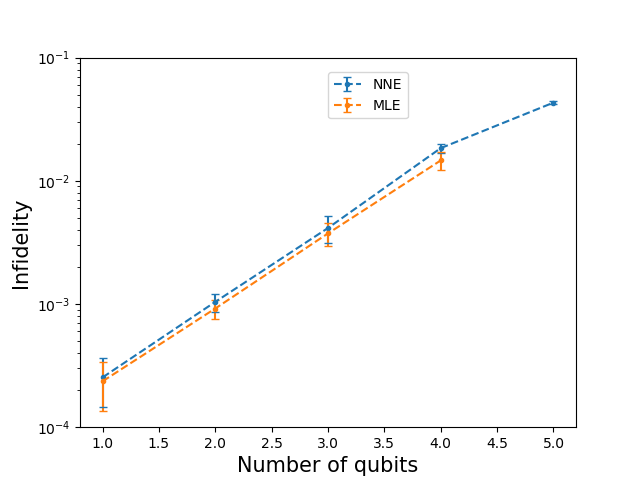}
    \caption{The infidelity of NNE and MLE on the maximally mixed state. $N_0 = 2000$}
    \label{fig:F-n_mixed_state}
\end{figure}

The maximally-mixed state often gives the larger mean square Hilbert-Schmidt (HS) distance than the other states \cite{hou2016full}. Fig.~\ref{fig:F-n_mixed_state} shows the infidelity of NNE and MLE on maximally mixed state with different qubit number and different copies of qubits used in numerical experiments.

We see that the performance of MLE and NNE are also comparable on the maximally mixed states with up to 4 qubits.

\section{\label{sec:level1}Computational complexity analysis}

\subsection{Computational complexity of a single state reconstruction}

\begin{figure}[h!]
\centering
\subfigure{
 \includegraphics[width = 0.45\textwidth ]{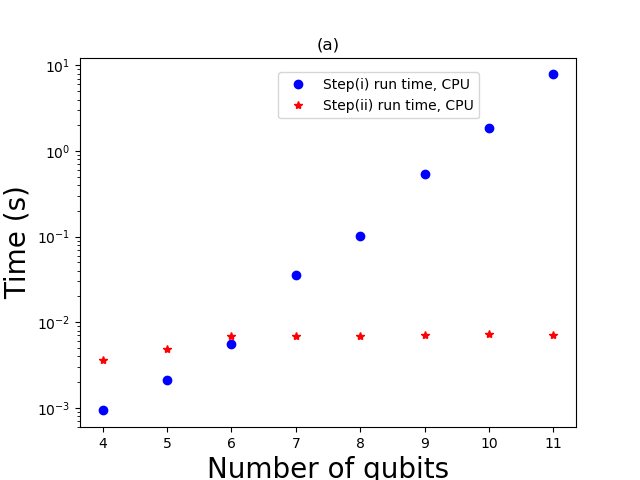}
 \label{fig:Step1-Step2}
 }
 \hspace{0.1in}
 \subfigure{
 \includegraphics[width = 0.45\textwidth ]{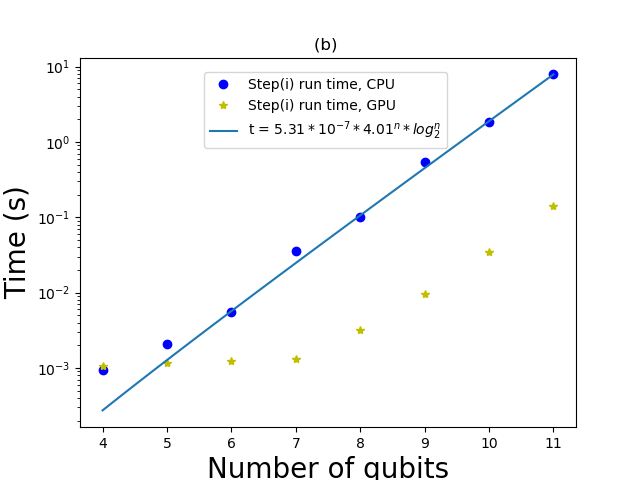}
 \label{fig:CPU-GPU}
 }
 \caption{(a) Actual run time of Step(i) and Step(ii) on CPU. (b) Actual run time of Step(i) on CPU and GPU}
 \label{fig:time_complexity}
\end{figure}

When the network is well trained, i.e. all the parameters are set, there is only a single feed-forward calculation in a state reconstruction process. For generality, we consider the informationally complete measurement set with minimum number of measurements, which is $d^2$. In this case, the number of neurons in the input layer is $d^2$ and $2d^2$ in the output layer. As for the number of neurons in two hidden layers, there is no precise mathematical prediction. In practice, we find a linear scale with qubit number $n$ is enough to produce accurate results. Therefore, the computational complexity in step(i) is $O(d^2 \log d)$. Step(ii) includes a multiplication of two $d*d$ matrices and has a basic complexity of $O(d^3)$. So the overall computational complexity of our model is $O(d^3)$, which is the fastest one among existing state estimation algorithms for FQST. And we can see that step(i), which is the core of our method, only has $O(d^2 \log d)$ computational complexity while step(ii), which aims to put physical constraint on the output matrix via a simple matrix multiplication, has the dominant time complexity. Actually faster matrix multiplication algorithms have been proposed, among which the fastest known one has a time complexity of $O(d^{2.373})$ \cite{le2014powers}. So if fastest matrix multiplication algorithms are applied the computational complexity of our model can further decrease.

Figs.~\ref{fig:Step1-Step2},~\ref{fig:CPU-GPU} numerically show the actual run time of a single state reconstruction process. The computer hardware includes a 64GB RAM, a Xeon E5-2660v4 CPU with 2 GHz, 8 cores and a Tesla P100 GPU.

Although step(ii) has larger computational complexity than step(i), its actual run time depends on different matrix multiplication algorithms and programming languages. And for relatively small systems step(ii) takes much smaller time than step(i). We use $\rm{tf.matmul}$ function within tensorflow framework to perform the matrix multiplication and as shown in Fig.~\ref{fig:Step1-Step2} the run time of step(ii) barely changes with the number of qubits in our case and the scale of the total computational complexity is dominated by step(i) for systems with up to 11 qubits.

In Fig.~\ref{fig:CPU-GPU} we fit the actual run time of step(i) with CPU by 
\begin{equation}
    t = A*x^n*\log n
\end{equation}
we get 
$$
A = 5.1*10^{-7};\ x = 4.01
$$
which is congruent with our theoretical prediction. For a 8-qubit system, our algorithm takes around 0.1s, in great contrast to MLE which takes one week \cite{da2011practical, haffner2005scalable}. And when GPU is used to support the parallel computation based on tensorflow framework we can get further dramatic acceleration. For a 10-qubit system it can be about 50 times faster. 

\begin{table}[h!]
    \centering
    \begin{tabular}{| p{4cm}<{\centering}|p{3cm}<{\centering}|}
        \hline
        State estimation model & Computational complexity \\
        \hline
        MLE & $>O(d^4)$ \\
        \hline
        Efficient MLE\cite{smolin2012efficient} & $O(d^4)$ \\
        \hline
        LRE\cite{qi2013quantum} & $O(d^4)$ \\
        \hline
        Bayesian mean estimation (BME)\cite{blume2010r} & $>O(d^4)$ \\
        \hline
        NNE & $O(d^3)$ \\
        \hline
    \end{tabular}
    \caption{Computational complexities for different state estimation models (There are no precise values for traditional MLE and BME.)}
    \label{tab:complexity}
\end{table}

Table,~\ref{tab:complexity} shows the efficiency comparison in terms of computational complexity between our model and other state estimation models for full quantum state tomography. We see our model is the fastest one.

\subsection{Computational complexity of the training process}
As the supervised learning is based on big data, the training process is usually time-consuming. So the computational complexity of this process depends on the scale of the training set. In machine learning, it is commonly accepted that the number of training examples in the training set should increase linearly with the dimension of the input feature vector. Thus there should be $O(d)$ training examples in our method and the computational complexity of the training process is $O(d^4)$.

Though the training process is time-consuming, it is not a vital problem. As this is a universal tomographic scheme, as long as one network is well trained and parameters are saved it only requires a single feed-forward calculation for new QST tasks when optimal parameters are loaded. So under this circumstance, the training process could be finished in advance with powerful computational resources.  

\section{Discussion and Conclusion}

An efficient state estimation model empowered by machine learning technology has been presented in this work. Unlike previous efforts in reducing the measurement number in tomography it focuses on speeding up the data analysis of full quantum state tomography, which assumes no prior knowledge of the state of a quantum system and is the most versatile case. Because our model is based on supervised learning, once the training on artificial data set is finished new state reconstruction does not include fitting or regression process. 

The computational complexity for a single state reconstruction process is $O(d^3)$, which is much faster than MLE and is the fastest one among existing full quantum state tomography algorithms for FQST (see Table,~\ref{tab:complexity}). Due to the limited computing resources, we currently can only train our model for systems with up to 5 qubits. It is prospective that our model can apply to larger systems from the time complexity point of view.

\begin{acknowledgments}
We would like to thank Sisi Zhou and Ryan Shaffer for helpful discussions.

\end{acknowledgments}

\appendix
\renewcommand{\appendixname}{Appendix}

\section{Generation of the training set}

An ensemble of general states of a quantum system is in general described by a probability prior. For our training purpose, the probability prior should be able to produce a uniform spread throughout the allowed state space. This uniformity is defined based on the Bures distance

\begin{equation}
    D_B(\rho_1,\rho_2)=\sqrt{2(1-Tr[(\sqrt{\rho_1}\rho_2\sqrt{\rho_1})^\frac{1}{2}])}.
\end{equation}
It corresponds to the minimal Fubini-Study distance between all possible purifications of both mixed states $\rho_1$ and $\rho_2$.\cite{zyczkowski2001induced} It also corresponds to maximal randomness of the states. The infinitesimal distance element between two states $\rho$ and $\rho+\delta\rho$ is
\begin{equation}
    (d_{S_B})^2=2\sum_{j,k}\frac{|\left\langle j|\delta\rho|k\right\rangle|^2}{\lambda_j+\lambda_k},
\end{equation}
where $\rho$ has orthonormal eigenvectors ${\left|\lambda_j\right\rangle}$ with eigenvalues $\lambda_j$. The volume element which is based on distance element is calculated as\cite{hall1998random}:
\begin{equation}
    dV_B=\frac{d\lambda_1...d\lambda_M}{(\lambda_1...\lambda_M)^\frac{1}{2}}\prod\limits_{i<j}4\frac{(\lambda_i-\lambda_j)^2}{\lambda_i+\lambda_j}d_{x_{ij}}d_{y_{ij}}
\end{equation}
Normalising $dV_B$ yields the desired probability distribution over the space of density operators. The distribution is written as
\begin{equation}
    P_B(\lambda_1,...,\lambda_N)=C^B_N\prod\limits_i\lambda_i^{-\frac{1}{2}}\prod\limits_{i<j}^{1...N}\frac{(\lambda_i-\lambda_j)^2}{\lambda_i+\lambda_j},
\end{equation}
where $C^B_N$ represents the normalization constant. It ensures that two different regions in state space that have the same volume occupied with the same probability. 

To generate a set of random density matrixes $\{\rho_t\}$ with respect to the Bures distribution, we proceed according to the following steps.

(i)Generate a complex random matrix $G$ of size $N$ pertaining to the Ginibre ensemble.

(ii)Generate a random unitary matrix $U$ distributed according to the Haar measure on $U(N)$.\cite{pozniak1998composed}

(iii)Output a random density matrix
\begin{equation}
    \rho_B=\frac{(1+U)GG^\dagger(1+U^\dagger)}{Tr[(1+U)GG^\dagger(1+U^\dagger)]}.
\end{equation}

\end{document}